\def\year{2019}\relax
\documentclass[letterpaper]{article} 
  \usepackage{aaai19}  
  \usepackage{times}  
  \usepackage{helvet}  
  \usepackage{courier}  
  \usepackage{url}  
  \usepackage{graphicx}  
  \usepackage{appendix}
\usepackage{url}            
\usepackage{amsfonts}       
\usepackage{nicefrac}       
\usepackage{microtype}      
\usepackage{enumitem}
\usepackage{float,subfig}
\usepackage[svgnames]{xcolor}
\usepackage{soul}
\usepackage[export]{adjustbox}
\usepackage{calc}

\usepackage{tikz}
\usetikzlibrary{positioning}

\usepackage{amsmath,amssymb,amsthm}
\usepackage{newtxtext}
\usepackage[mathscr]{eucal}
\usepackage{stmaryrd}

\usepackage{upgreek}
\newcommand{\citet}[1]
{\citeauthor{#1} ̃\shortcite{#1}}
\newcommand{\citep}{\cite}

\usepackage{colortbl}
\usepackage{multirow}
\usepackage{algorithm}
\usepackage[noend]{algpseudocode}

\usepackage{pgfplots}
\usepgfplotslibrary{dateplot}
\usepackage{anyfontsize}

\usepackage{pgfcalendar}
\usepackage{xargs}

\usepackage[mathscr]{eucal}
\usepackage{lipsum}  

\definecolor{navy}{rgb}{0.1, 0.1, 0.8}
\definecolor{gray}{rgb}{0.4, 0.4, 0.4}
\definecolor{olive}{rgb}{0.1, 0.5, 0.1}
\definecolor{ruby}{rgb}{0.8, 0.1, 0.3}

\newcommand{\x}{\mathbf{x}}
\newcommand{\y}{\mathbf{y}}

\newcommand{\bX}{\mathbf{X}}

\newcommand{\FCal}{\mathscr{F}}

\newcommand{\UCal}{\mathscr{U}}

\newcommandx{\todom}[2][1=]{\todo[linecolor=blue,backgroundcolor=blue!25,bordercolor=blue,#1]{#2}}
\newcommandx{\revisit}[2][1=]{\todo[linecolor=Plum,backgroundcolor=Plum!25,bordercolor=Plum,#1]{#2}}
\newcommand{\redo}[1]{{#1}}

\nocopyright
\begin{document} 
\title{Comparative Document Summarisation via Classification}

\author{Umanga Bista$^{\star\ddag}$\ Alexander Mathews$^{\star\ddag}$\ Minjeong Shin$^{\star\ddag}$\ Aditya Krishna Menon$^{\star}$\thanks{Now at Google Research.}\ Lexing Xie$^{\star\ddag}$\\
Australian National University$^{\star}$, Data to Decision CRC $^{\ddag}$\\
{\tt\{umanga.bista,alex.mathews,minjeong.shin,aditya.menon,lexing.xie\}@anu.edu.au}
}
\maketitle

\begin{abstract}


This paper considers extractive summarisation in a \emph{comparative} setting:
given two or more document groups (e.g., separated by publication time),
the goal is to select a small number of documents that are representative of each group, 
and also maximally distinguishable from other groups.
We formulate a set of new objective functions for this problem that 
connect recent literature on 
document summarisation, interpretable machine learning, and data subset selection. 
In particular, 
by casting the problem as a binary classification amongst different groups,
we derive objectives
based on the notion of maximum mean discrepancy, as well as a
simple yet effective gradient-based optimisation strategy. 
Our new formulation allows scalable evaluations of comparative summarisation as a classification task,
both automatically and via crowd-sourcing.
To this end,
we evaluate comparative summarisation methods on a newly curated collection 
of controversial news topics over 13 months. 
We observe that gradient-based optimisation outperforms discrete and baseline approaches 
in \redo{14} out of 24 different automatic evaluation settings. 
In crowd-sourced evaluations, summaries from gradient optimisation 
elicit 7\% more accurate classification 
from human workers 
than discrete optimisation. 
{Our result contrasts with recent literature on 
submodular data subset selection that favours discrete optimisation.}
{We posit that our formulation of comparative summarisation will prove 
useful 
in a diverse range of use cases such as comparing content sources, authors, related topics, or distinct view points.
}


\end{abstract}

\section{Introduction} 

\begin{figure}
    \centering
   \includegraphics[width=0.9\linewidth]{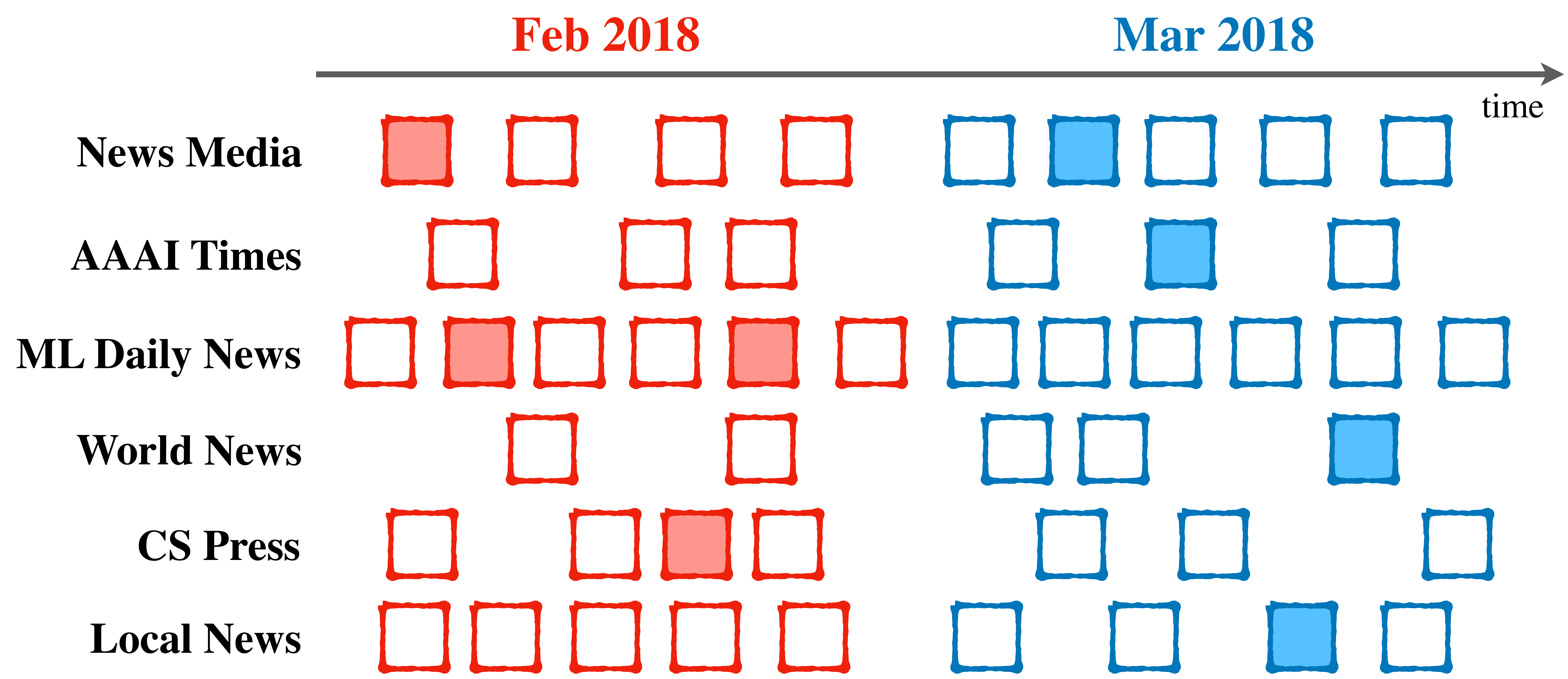}
    \caption{An illustrative example of comparative summarisation. Squares are news articles, rows denote different news outlets, and the $x$-axis denotes time. The shaded articles are chosen to represent AI-related news during {Feb} and {March} 2018, respectively. They aim to summarise topics in each month, and also highlight differences \emph{between} the two months.}
    \label{fig:news}
\end{figure}

Extractive summarisation is the task of selecting a few representative documents from
a larger collection.
In this paper, we consider {\em comparative summarisation}:
given \emph{groups} of document collections, 
the aim is to
select documents that represent each group,
but also highlight differences \emph{between} groups.
This is in contrast to traditional document summaries which aim to {represent each group by independently optimising for coverage and diversity, without considering other groups}.
As a concrete example, given thousands of news articles per month on a certain topic, 
groups can be formed by publication time, by source,
or by political leaning. 
{Comparative} summarisation systems can then help answer user questions 
such as: what is new on the topic of climate change this week, 
what is different between the coverage in NYTimes and BBC, 
or what are the key articles covering the carbon tax and the Paris agreement?
{In this work, we focus on highlighting changes within a long running 
news topic over time;
see Figure~\ref{fig:news} for an illustration.}


Existing methods for extractive summarisation use a variety of formulations such as structured prediction~\citep{Li:2009:EDC:1526709.1526720}, 
optimisation of submodular functions~\citep{Lin:2011:CSF:2002472.2002537}, 
dataset interpretability~\citep{kim2016examples}, and
dataset selection via submodular optimisation~\citep{mirzasoleiman2016fast,wei2015submodularity,mitrovic2018data}. 
Moreover, recent formulations of comparative summarisation use discriminative sentence selection~\citep{wang2012comparative,li2012multi}, or highlight differences in common concepts across documents ~\citep{huang2011comparative}.
But the connections and distinctions of these approaches has yet to be clearly articulated. 
To evaluate summaries, traditional approaches employ automatic metrics such as {\sf ROUGE} ~\citep{lin2004rouge} on manually constructed summaries~\citep{lin2003automatic,nenkova2007pyramid}. 
This is difficult to employ for new tasks and new datasets, and does not scale. 

Our approach to comparative summarisation is based on a novel formulation of the problem
in terms of two \emph{competing classification tasks}.
Specifically, we formulate the problem 
as finding
summaries for each group such that a powerful classifier can distinguish them from documents belonging to \emph{other} groups,
but cannot distinguish them from documents belonging to the \emph{same} group.
%
We show how this framework encompasses an existing nearest neighbour objective for summarisation,
and
propose two new objectives based on the maximum mean discrepancy~\cite{gretton2012kernel} -- 
{\em mmd-diff} which emphasises classification accuracy and {\em mmd-div} which emphasises summary diversity 
--
as well as 
new gradient-based optimisation strategies for 
these objectives.
%

A key advantage of 
our discriminative problem setting
is that it 
allows summarisation to be evaluated as a classification task.
To this end,
we design automatic and crowd-sourced evaluations {for comparative summaries, which we apply on a new dataset of three ongoing controversial news topics.}
We observe that the new objectives with gradient optimisation are top-performing
in \redo{14} out of 24 settings (across news topics, summary size, and classifiers)
(\S \ref{ssec::auto_eval}).
We design a new crowd-sourced article classification task for human evaluation. 
We find that workers are on average 7\% more accurate in classifying articles 
using summaries generated by {\em mmd-diff} with gradient-based optimisation than 
all alternatives. Interestingly, our results 
contrast with the body of work on dataset selection and summarisation 
that favour discrete greedy optimisation of submodular objectives due to approximation guarantees.
We hypothesise that the comparative summarisation problem is particularly amenable to 
gradient-based optimisation due to the small number of prototypes needed.
Moreover, gradient-based approaches can further improve solutions found by greedy approaches.

In sum, the main contributions of this work are:
\begin{itemize}
	\item A new formulation of comparative document summarisation in terms of competing binary classifiers, two new objectives based on this formulation, and their corresponding gradient-based optimisation strategies.
	\item Design of a scalable automatic and human evaluation methodology for comparative summarisation models, with results showing that the {new objectives out-perform existing submodular objectives.} 
    \item A use case of comparatively summarising articles over time from a news topic on a new dataset\footnote{Code, datasets and a supplementary appendix are available at https://github.com/computationalmedia/compsumm} of three controversial news topics from 2017 to 2018.
\end{itemize}

\section{Related Works}
The broader context of this work is extractive summarisation. 
Approaches to this problem include incorporating diversity measures 
from information-retrieval~\citep{Carbonell:1998:UMD:290941.291025}, 
structured SVM regularised by constraints for diversity, 
coverage, and balance~\citep{Li:2009:EDC:1526709.1526720}, 
or topic models for summarisation~\citep{Haghighi:2009:ECM:1620754.1620807}. 
Time-aware summarisation is an emerging subproblem, 
where the current focus is on modeling continuity~\citep{ren2016time} or 
continuously updating summaries~\citep{RuckleG17}, 
rather than formulating comparisons.
~\cite{li2012multi,wang2012comparative} present methods to 
extract one or few discriminative sentences from a small multi-document corpus 
utilising greedy optimisation and evaluating qualitatively. 
~\citep{huang2011comparative} compares descriptions about similar concepts 
in closely related document pairs, leveraging an integer linear program and evaluating 
with few manually created ground truth summaries. 
While these works exist in the  domain of comparative summarisation, they are either specific to a data domain or have
evaluations which are hard to scale up.
In this paper we present approaches to comparative summarisation with 
intuition from competing binary classifiers, leading to different objectives 
and evaluation. We demonstrate and evaluate the application of these approaches 
to multiple data domains such as images and text.

Submodular functions have been the preferred form of discrete objectives for summarising text~\citep{Lin:2011:CSF:2002472.2002537}, images~\citep{simon2007scene} and data subset selection~\citep{wei2015submodularity,mitrovic2018data}, 
since they can be optimised greedily with tightly-bounded guarantees. 
The topic of interpreting dataset and models use similar strategies~\citep{kim2016examples,bien2011prototype}.
This work re-investigates classic continuous optimisation for comparative summarisation, 
and puts it back on the map as a competitive strategy. 
\section{Comparative Summarisation as Classification}
\label{sec:method}

Formally, the comparative summarisation problem is defined on
$G$ groups of document collections $\{ \bX_1, \ldots, \mathbf{X}_G\}$, 
where a group may, for example, correspond to news articles about a specific topic published in a certain month.
We write the document collection for group $g$ as $$ \mathbf{X}_g =
\{\x_{g,1}, \x_{g,2}, \ldots , \x_{g,N_g}\}$$ where
$N_g$ is the total number of documents in group $g$. We represent individual
documents as vector $\x_{g, i} \in \mathbb{R}^d$ (see \S\ref{sec:exp}).

\begin{figure*}
    \centering
    \subfloat[Coverage is 
    the average similarity between 
    the test prototype document and all other documents in the test documents's 
    group.]{
        \includegraphics[width=0.31\linewidth]{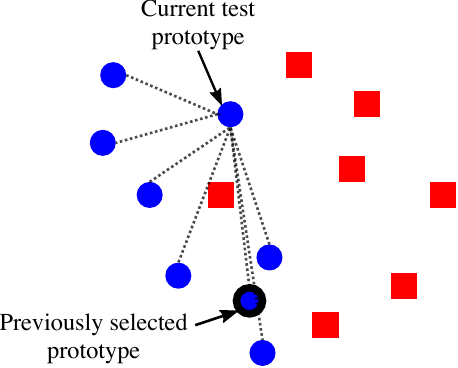}
        \label{fig:rep}
    } \quad
    \subfloat[Discriminativeness is 
    the average similarity between 
    the test prototype document and all documents \textbf{not} in the test document's 
    group.]{
        \includegraphics[width=0.31\linewidth]{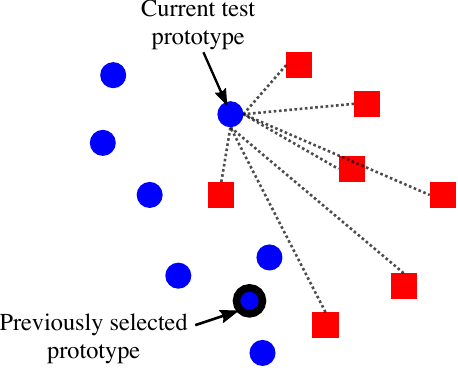}
        \label{fig:desc}
    }\quad
    \subfloat[Diversity is 
        the average dissimilarity between 
        the test prototype document and all selected prototypes in the test 
        document's group.]{
        \includegraphics[width=0.31\linewidth]{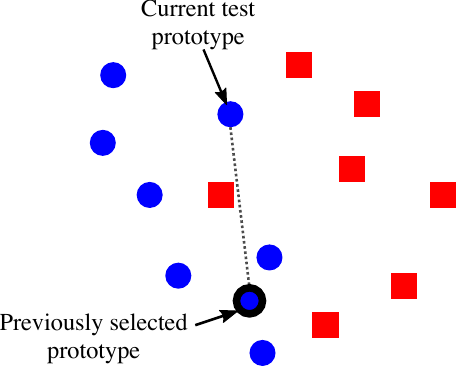}
        \label{fig:div}
    }
    \caption{Illustration of coverage, discriminativeness and diversity 
    criteria for selecting prototypes. The two document groups are shown as
    blue circles and red squares. The dotted lines represent comparisons 
    between pairs of documents.}
\end{figure*}

Our goal is to summarise each document collection $\mathbf{X}_g$ with a set of
\emph{summary documents} or \emph{prototypes} $\bar{\mathbf{X}}_{g} \subset \mathbf{X}_{g}$, written $$
\bar{\mathbf{X}}_{g} = \{ \bar{\x}_{g,1}, \bar{\x}_{g,2},
\ldots, \bar{\x}_{g, M} \}$$ For simplicity, we assume the
number of prototypes $M$ is the same for each group.
The selected
prototypes should represent the documents in the group achieving coverage  (Figure~\ref{fig:rep}) and diversity (Figure~\ref{fig:div}), while simultaneously
discriminating documents from other groups (Figure~\ref{fig:desc}). 
{For example, if we have news articles on the \emph{Climate Change} topic 
then they may discuss the \emph{paris agreement} in February,  
\emph{coral bleaching} in March, and \emph{rising sea levels}
in both months. A comparative summary should include 
documents about the \emph{paris agreement} in February and \emph{coral 
bleaching} in March, but potentially not on \emph{rising sea levels} as 
they are common to both time ranges and hence do not discriminate.}

 


\subsection{A Binary Classification Perspective}
\label{ssec:classify_perspective}

We now cast comparative summarisation as a binary classification problem.
To do so, let us re-interpret the two defining characteristics of prototypes $\bar{\mathbf{X}}_g$ for the $g$th group:
\begin{enumerate}[label=(\roman*),leftmargin=\widthof{[ii]}+\labelsep]
    \item they must represent the documents belonging to that group.
Intuitively, this means that each $\bar{\x}_{g, i} \in \bar{\mathbf{X}}_{g}$ must be \emph{indistinguishable} from all
$\x_{g, j} \in \mathbf{X}_{g}$.

    \item they must discriminate against documents from all other groups.
Intuitively, this means that each $\bar{\x}_{g, i} \in \bar{\mathbf{X}}_{g}$ must be \emph{distinguishable} from
$\x_{\lnot g, j} \in \mathbf{X}_{\lnot g}$, 
where $\mathbf{X}_{\lnot g}$ denotes the set of all documents belonging to all groups except $g$.
\end{enumerate}

This lets us relate prototype selection to the familiar binary classification problem:
for a good set of prototypes,
\begin{enumerate}[label=(\alph*),leftmargin=\widthof{[b]}+\labelsep]
    \item there \emph{cannot exist} a classifier that can accurately discriminate between them and documents from that group.
For example, even a powerful classifier should not be able to discriminate 
prototype documents about the Great Barrier Reef from other documents about the Great Barrier Reef.

    \item there \emph{must exist} a classifier that can accurately discriminate them against documents from all other groups.
For example, a reasonable classifier should be able to discriminate prototypes about the Great Barrier Reef from documents about emission targets. 
\end{enumerate}
Consequently, we can think of prototype selection in terms of two competing binary classification objectives:
one distinguishing $\bar{\mathbf{X}}_g$ from $\mathbf{X}_g$,
and another distinguishing $\bar{\mathbf{X}}_g$ from $\mathbf{X}_{\lnot g}$.
In abstract, this suggests a multi-objective optimisation problem of the form 
\begin{equation}
    \max_{\bar{\mathbf{X}}_1, \ldots, \bar{\mathbf{X}}_G}
    \left( \sum_{g = 1}^{G} -\mathrm{Acc}( \bar{\mathbf{X}}_g, \mathbf{X}_{g} ), \sum_{g = 1}^{G} \mathrm{Acc}( \bar{\mathbf{X}}_g, \mathbf{X}_{\lnot g}) \right),
    \label{eqn:multi-obj}
\end{equation}
where $\mathrm{Acc}( \mathbf{X}, \mathbf{Y} )$ estimates the accuracy of the best possible classifier for distinguishing between the datasets $\mathbf{X}$ and $\mathbf{Y}$.
Making this idea practical requires committing to a particular means of balancing the two competing objectives.
More interestingly, one also needs to find a tractable way to estimate $\mathrm{Acc}( \cdot, \cdot )$:
explicitly searching over rich classifiers such as deep neural networks, would lead to a computationally challenging nested optimisation problem.

In the following we discuss a set of objective functions that avoid such nested optimisation.
We also discuss two simple optimisation strategies for these objectives in \S\ref{sec:opt}.



\subsection{Prototype Selection via Nearest-neighbour }

One existing prototype selection method involves approximating the 
intragroup 
$\mathrm{Acc}( \cdot, \cdot 
)$ term
in Eq~\ref{eqn:multi-obj} using nearest-neighbour classifiers, while ignoring 
the intergroup accuracy term. 
Specifically, a
formulation of prototype selection
in ~\cite{wei2015submodularity} 
maximises the total similarity of every point to its nearest prototype from the same class:
\begin{equation}
  \UCal_{\text{\em nn}}(\bar\bX) = \sum_{g=1}^G\sum_{i=1}^{N_g} \max_{m \in \{ 1, \ldots, M \}} \mathrm{Sim}(\bar\x_{g,m}, \x_{g,i})
  \label{eq:nncomp}
\end{equation}
Here, 
$\mathrm{Sim}$
is any similarity function, with admissible choices including a negative distance, or valid kernel functions.

The nearest neighbour utility function is simple and intuitive.
However, it only
considers the most similar prototype for each datapoint which
misses our second desirable property of prototypes: that they explicitly distinguish between different classes. Moreover, the nearest neighbour utility function can be challenging to optimise because of the $\max$ function. The rest of this section introduces
three other utilities that address these concerns.

\subsection{Preliminaries: Maximum Mean Discrepancy}

The \emph{maximum mean discrepancy} (\emph{MMD})~\citep{gretton2012kernel}
measures the distance between two distributions
by leveraging the kernel trick~\citep{Scholkopf:2002}.
Intuitively, MMD deems 
two distributions
to be close if
the \emph{mean} of \emph{every} function in some rich class $\FCal$ is close under both distributions.
For suitable $\FCal$, this is equivalent to comparing the moments of the two distributions;
however, a na\"{i}ve implementation of this idea would require a prohibitive number of evaluations.
Fortunately, choosing $\FCal$ to be a reproducing kernel Hilbert space (RKHS)
with kernel function $k(\cdot, \cdot)$ leads to an expression that is defined
only in terms of document interactions via the kernel function~\citep{gretton2012kernel}:
\begin{align}
    \mathrm{MMD}^2(\mathbf{X}, \mathbf{Y}) =\, &\mathbb{E}_{\x,\x'} [k(\x,\x')] - 2 \cdot \mathbb{E}_{\x,\mathbf{y}} [k(\x,\mathbf{y})] + \nonumber \\
    &\mathbb{E}_{\mathbf{y},\mathbf{y}'} [k(\mathbf{y},\mathbf{y}') ]  
    \label{equ:mmd2}
\end{align}
where $\x \sim \mathbf{X}, \mathbf{y} \sim \mathbf{Y}$ are observations
from two datasets $\mathbf{X}, \mathbf{Y}$.
In practice, it is common to use the radial basis function (RBF) or
Gaussian kernel $k( \x, \mathbf{y} ) = e^{-\gamma \cdot \| \x - \mathbf{y} \|_2^2}$
with fixed bandwidth $\gamma > 0$.

One often approximates MMD using sample expectations:
given $n$ samples $\x_1,\ldots,\x_n$ from $\mathbf{X}$, and $m$ samples $\y_1,\ldots,\y_n$ from $\mathbf{Y}$,
we may compute
\begin{align}
   & \mathrm{MMD}^2(\mathbf{X}, \mathbf{Y}) =
   \frac{1}{n^2} \sum_{i=1}^{n}\sum_{j=1}^{n} k(\x_i,\x_j) \nonumber \\
    & - \frac{2}{mn} \sum_{i=1}^{n}\sum_{j=1}^{m} k(\x_i,\y_j) +
    \frac{1}{m^2} \sum_{i=1}^{m}\sum_{j=1}^{m} k(\y_i,\y_j)
    \label{equ:mmd3}
\end{align}

%
\subsection{Prototype Selection via MMD}

One can think of MMD as {implicitly} computing a (kernelised)
\emph{nearest centroid classifier} to distinguish between $\bX$ and $\mathbf{Y}$:
MMD is small when this classifier has high expected error.
Thus, MMD can be seen as an efficient approximation to classification
accuracy $\mathrm{Acc}( \cdot, \cdot )$.
This intuition lead to a practical utility function that approximates
Equation~\ref{eqn:multi-obj}
by taking the difference of two MMD terms:
\begin{equation}
     \resizebox{.9\hsize}{!}{$
     \UCal_{\text{\em diff}}(\bar\bX) = \sum_{g}
     ( -\mathrm{MMD}^2(\bar{\mathbf{X}}_g, \mathbf{X}_g) + \lambda \cdot \mathrm{MMD}^2(\bar{\mathbf{X}}_g, \mathbf{X}_{\lnot g})) 
    $}
    \label{equ:mmd_labels}
\end{equation}

The hyper-parameter $\lambda$ trades off how well the prototype represents its group, against how well it distinguishes between groups (Figure~\ref{fig:desc}).
Intuitively, when the term $\mathrm{MMD}^2( \bar{\mathbf{X}}_g, \mathbf{X}_{\lnot g})$ is large then the prototypes $\bar{\mathbf{X}}_g$ are dissimilar from documents $\mathbf{X}_{\lnot g}$ of other groups.
Similarly, when $\mathrm{MMD}^2(\bar{\mathbf{X}}_g, \mathbf{X}_{g})$ is small then the prototypes are similar to documents of that group.
Maximising $-\mathrm{MMD}^2$ gives prototypes that are both close to the empirical samples 
(as seen by the $\mathbb{E}_{\x,\mathbf{y}}$ term in Equation~\ref{equ:mmd2} and 
illustrated by Figure~\ref{fig:rep}) and far from one another (as seen by the
 $\mathbb{E}_{\mathbf{y},\mathbf{y}'}$ term and illustrated by Figure~\ref{fig:div}).

While the objective of Equation~\ref{equ:mmd_labels} provides the core of our approach, we also present a variant that 
increases the diversity of prototypes chosen for each group.
{A closer examination of the difference of $\mathrm{MMD}^2$ in Equation~\ref{equ:mmd_labels}
-- by expanding both using Equation~\ref{equ:mmd2} --
reveals two separate prototype diversity terms
$-\mathbb{E}_{\bar\x_g,\bar\x_g'}[k(\bar{\x}_g,\bar\x_g') ]$
and $\lambda \mathbb{E}_{\bar\x_g,\bar\x_g'}[k(\bar\x_g,\bar\x_g') ]$.
The latter counteracts the former and decreases prototype diversity (details in Appendix$^1$).}
On the expanded form of $\lambda\mathrm{MMD}^2(\bar{\mathbf{X}}_g, \mathbf{X}_{\lnot g})$, we remove 
the terms not involving $\bar{\x_g}$, as they are constants and have no effect on the solution, and
also remove the conflicting diversity term $\lambda \mathbb{E}_{\bar\x_g,\bar\x_g'}[k(\bar\x_g,\bar\x_g') ]$. This gives
a new objective:
\begin{equation}
        \resizebox{.9\hsize}{!}{$
\UCal_{\text{\em div}}(\bar\bX) = \sum_{g} ( -\mathrm{MMD}^2(\bar{\mathbf{X}}_g, \mathbf{X}_g)
 - 2 \lambda \mathbb{E}_{\bar\x_g, \x_{\lnot g}} [k(\bar\x_g, \x_{\lnot g})]  )
$}
 \label{equ:mmd_labels_div}
\end{equation}


Maximising $-\lambda \mathbb{E}_{\bar\x_g, \x_{\lnot g}} [k(\bar\x_g, \x_{\lnot g})]$ encourages prototypes in group $g$ to be far from data points in other groups. 




One can envision another variant that explicitly optimises the diversity between
{\em prototypes of different classes}, rather than between prototypes of class $g$
against data points in other classes. This is computationally more efficient,
and reflects similar intuitions.
However, it did not outperform $\UCal_{\text{\em diff}} $, $\UCal_{div}$ in summarisation tasks, and is omitted due to space limitations.

\noindent{\bf{Differences to related objectives.}}
The nearest-neighbour objective was articulated in~\cite{wei2015submodularity} and earlier in~\cite{bien2011prototype}, and used for classification tasks.
Recently, ~\cite{kim2016examples} proposed {\em MMD-critic}, 
which selects prototypes $\bar{\mathbf{X}}$ for a single group of documents $\bX$ by maximizing 
$ -\mathrm{MMD}^2(\bar\bX, \bX). $
The first term in Equation \ref{equ:mmd_labels} builds on this formulation, applying this idea independently for each group.
Our second term is crucial to encourage prototypes that \emph{only} represent their own group
and none of the other groups.
{\em MMD-critic} also contains {\em model criticisms}, 
which have to be optimized sequentially after obtaining prototypes. As shown in \S\ref{sec:exp}, {\em MMD-critic}
under-performs in comparison tasks by a significant margin.

\section{Optimising Utility Functions}
\label{sec:opt}

{There are two general strategies for optimising the utility functions
outlined in \S\ref{sec:method} to generate summaries that are
a subset of the original dataset:
greedy and gradient optimisation.}

\noindent{\bf Greedy optimisation}.
The first strategy involves directly choosing $M$ prototypes for each group. 
Obtaining the exact solution to this discrete optimisation problem is intractable; 
however, approximations such as greedy selection can work well in practice,
and may also have theoretical guarantees.

Specifically, suppose we wish to maximise a utility set function $F : 2^{|V|} \to \mathbb{R}$ defined on ground set $V$.
For $S \subset V$ and $s \in V \setminus S$, the marginal gain of adding element $s$ to an existing set $S$ is known as the discrete derivative, and is defined by $\Delta_F(s| S) = F(S \cup s) - F(S)$.
We say $F$ is monotone if and only if the discrete derivatives are non-negative, i.e. $\Delta_F(s | S) \geq 0$, and is submodular if and only if the marginal gain satisfies diminishing returns, i.e. for $S \subseteq T \subset V, s \in V \setminus T$, $\Delta_F(s | S) \geq \Delta_F(s | T)$.
~\cite{nemhauser1978analysis} showed that if $F$ is submodular and monotone, greedy maximisation of $F$ yields an approximate solution no worse than $1-\frac{1}{e} \approx 0.63$ of the optimal solution under cardinality and matroid constraints.
~\cite{lin2010multi} showed this approximation holds with high probability even for non-monotone submodular objectives. 



%


In our context, given a utility function $\UCal$, the greedy algorithm
(see Appendix$^1$)
works by iteratively picking the $\x_g$ that provides the largest marginal gain ($\Delta_{\UCal}(\x_g | \bar{X}_g)$) one at a time for each group. 
Among the utility functions mentioned in \S\ref{sec:method}, the nearest-neighbour objective $\UCal_{nn}$ is submodular-monotone~\citep{wei2015submodularity}. 
The MMD function in Equation~\ref{equ:mmd2} is submodular-monotone under mild assumptions on the kernel matrix~\citep{kim2016examples}.
The MMD objective $\UCal_{\text{\em diff}}$ is the difference between 
two submodular-monotone functions,
which is not submodular in general. 
On the other hand,
the second term in $\UCal_{\text{\em div}}$ is modular with respect to $\bar\bX_g$, 
when the number of prototypes $M$ fixed and known in advance. 
Therefore, the diversity objective $\UCal_{\text{\em div}}$ is the difference between a submodular function and a modular function, and thus submodular.  
\noindent{\bf Gradient optimisation}
{The second strategy is 
to re-cast the problem to allow for continuous optimisation in the feature space,
e.g. using standard gradient descent.
To generate prototypes,
the solutions to this optimisation can then be {\em snapped} to the nearest data points as a post-processing step.}



Concretely, rather than searching for optimal prototypes $\bar{\mathbf{X}}_g$ directly,
we seek ``meta-prototypes''
$\bar{\mathbf{A}}_{g} = \{ \bar{\mathbf{a}}_{g,1}, \ldots, \bar{\mathbf{a}}_{g,M} \}$, drawn from the same space as the document embeddings. 
{We now modify $\UCal_{\text{\em diff}}$ (Equation~\ref{equ:mmd_labels}) to incorporate ``meta-prototypes''.} 
Note that $\UCal_{\text{\em div}}$ can be similarly modified, 
 but $\UCal_{\text{\em nn}}$ cannot, since the $\max$ function is not differentiable. 
The ``meta-prototypes'' for $\UCal_{\text{\em diff}}$ are chosen to optimise
\begin{equation}
    \footnotesize
    \max_{\bar{\mathbf{A}}_1, \ldots, \bar{\mathbf{A}}_G} \sum_{g} 
    ( -\mathrm{MMD}^2(\bar{\mathbf{A}}_g, \mathbf{X}_g) + 
    \lambda \cdot \mathrm{MMD}^2(\bar{\mathbf{A}}_g, \mathbf{X}_{\lnot g})  )
    \label{eqn:mmd-grad-obj}
\end{equation}
The only difference to Equation~\ref{equ:mmd_labels} is that we do \emph{not} enforce that $\bar{\mathbf{A}}_{g} \subset \mathbf{X}_g$.
{This subtle, but significant, difference allows Equation~\ref{eqn:mmd-grad-obj} 
to be optimized using gradient-following methods.} We use L-BFGS~\citep{byrd1995limited} 
with analytical gradients found in online appendix$^1$.
The selected meta-prototypes $\bar{\mathbf{A}}_g$ are then snapped to the nearest document in the group:
to construct the $i$th prototype for the $g$th group, we find
\begin{equation}
\bar{\x}_{g,i} = \underset{\x_{g,j} \in \mathbf{X}_g}{\mathrm{argmin}} \, \| \bar{\mathbf{a}}_{g, i} - \x_{g,j} \|_2^2.
\label{equ:snap_point}
\end{equation}
{On a problem often tackled with discrete greedy optimisation, one may wonder if gradient-based methods can be competitive;
we answer this in the affirmative in our experiments.}







\section{Datasets on Controversial News Topics}
\label{sec:datasets}

Exploring the evolution of controversial news topics
is a natural application of comparative summarisation. Comparative summarisation could help to better understand the role of news media in such a setting.
Recent work on controversial topics~\citep{garimella2018quantifying} focused on
the social network and interaction around controversial topics,{ but did not explicitly consider the content of news articles on these topics.}
To this end, we curate a set of news articles on long-running controversial topics
using tweets which link to news articles.
{We choose several long-running controversial topics with significant news coverage in 2017 and 2018.} To find articles relevant to these topics we use keywords to filter the Twitter stream, and adopt a snowball strategy to add additional keywords~\citep{verkamp2013five}. 
The articles linked in these tweets are then de-duplicated and filtered for spam. Article timestamps correspond to the creation time of the first tweet linking to it. Full details of the data collection procedure are described in online appendix$^1$.

\begin{figure}
  \begin{tikzpicture}\scriptsize
    \begin{axis}[date coordinates in=x,date ZERO=2017-06-01,
      xticklabel=\year-\month-\day, ymin=0, ymax=500, ytick={0,200,400},
      xtick={2017-08-01, 2018-01-01, 2018-06-01},
      ylabel style={align=center}, ylabel=\emph{Beef Ban}\\,
      height=2.4cm, width=0.98\linewidth,
      xmin=2017-06-01, xmax=2018-07-01]
      \addplot coordinates {
(2017-06-01, 131)
(2017-07-01, 454)
(2017-08-01, 222)
(2017-09-01, 81)
(2017-10-01, 108)
(2017-11-01, 65)
(2017-12-01, 56)
(2018-01-01, 75)
(2018-02-01, 46)
(2018-03-01, 41)
(2018-04-01, 57)
(2018-05-01, 45)
(2018-06-01, 93)
(2018-07-01, 69)
     };
    \end{axis}
  \end{tikzpicture}
  \begin{tikzpicture}\scriptsize
    \begin{axis}[date coordinates in=x,date ZERO=2017-06-23,
      xticklabel=\year-\month-\day, ymin=0, ymax=850, ytick={0,400,800},
      xtick={2017-07-14, 2017-08-25, 2017-10-07},
      max space between ticks=70, ylabel style={align=center}, ylabel=\emph{Capital}\\ \emph{Punish.},
      height=2.4cm, width=0.98\linewidth,
      xmin=2017-06-23, xmax=2017-10-21]
      \addplot coordinates {
(2017-06-23, 494)
(2017-06-30, 609)
(2017-07-07, 550)
(2017-07-14, 549)
(2017-07-21, 538)
(2017-07-28, 71)
(2017-08-04, 420)
(2017-08-11, 404)
(2017-08-18, 676)
(2017-08-25, 450)
(2017-09-01, 259)
(2017-09-08, 262)
(2017-09-15, 557)
(2017-09-23, 380)
(2017-09-30, 317)
(2017-10-07, 813)
(2017-10-14, 392)
(2017-10-21, 164)
     };
    \end{axis}
  \end{tikzpicture}
  \begin{tikzpicture}\scriptsize
    \begin{axis}[date coordinates in=x,date ZERO=2017-06-23,
      xticklabel=\year-\month-\day, ymin=0, ymax=2000, ytick={0,1000,2000},
      xtick={2017-06-30, 2017-07-14, 2017-07-28},
      ylabel style={align=center}, ylabel=\emph{Gun}\\ \emph{Control},
      height=2.4cm, width=0.98\linewidth,
      xmin=2017-06-23, xmax=2017-08-04]
      \addplot coordinates {
(2017-06-23, 757)
(2017-06-30, 1888)
(2017-07-07, 974)
(2017-07-14, 895)
(2017-07-21, 699)
(2017-07-28, 843)
(2017-08-04, 438)
     };
    \end{axis}
  \end{tikzpicture}
  \caption{Data volume over time for each topic}
  \label{figure::volumetimeplot}
\end{figure}
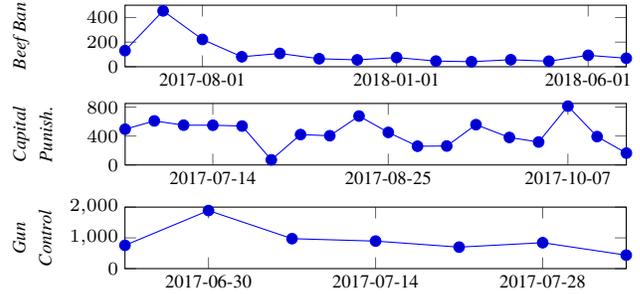

{In this work, we use news articles on three topics that appeared in a 14 month period (June 2017 -- July 2018). Within each topic we comparatively summarise news articles in different time periods to identify what has changed in that topic between the summarisation periods.}
To ensure our method works on a range of topics we chose substantially 
different long running topics: \textit{Beef Ban} -- controversy over the slaughter and 
sale of beef on religious grounds 
(1543 articles) is localised to a particular region, mainly Indian subcontinent, 
while \textit{Gun Control} -- restrictions on carrying, using, or purchasing firearms (6494 articles) and 
\textit{Capital Punishment} -- use of the death penalty (7905 articles) 
are topical in various regions around the world. 
Figure~\ref{figure::volumetimeplot} shows the number of new articles on each topic over time.


\begin{figure*}
    \centering
     \includegraphics[width=0.96\linewidth]{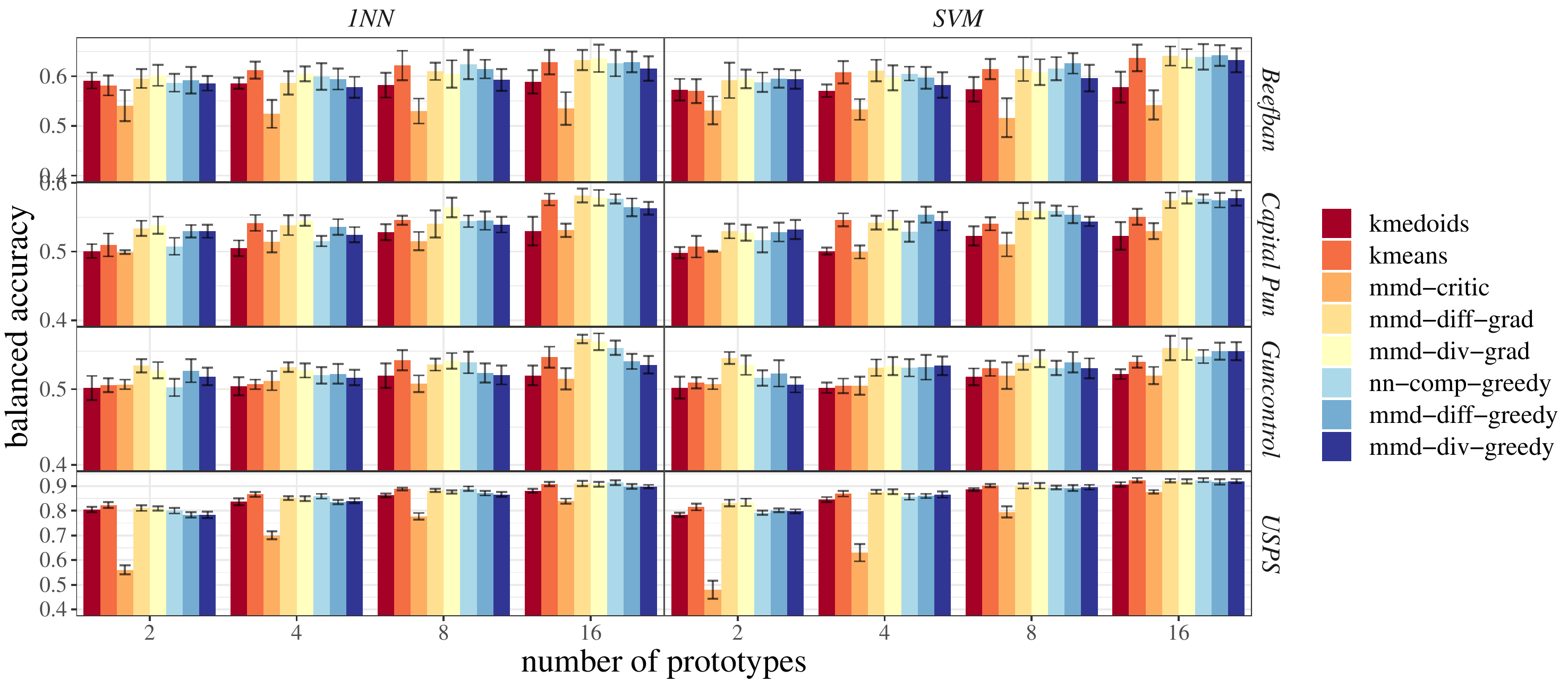}
    \caption{Comparative summarisation methods evaluated using the balanced 
    accuracy of 1-NN (left) and SVM (right) classifiers.
    Each row represent a dataset. Error bars show 95\% confidence intervals.}
    \label{fig:auto_bar}
\end{figure*}
\section{Experiments and Results}
\label{sec:exp}

We evaluate approaches to comparative summarisation using both automatic and crowd-sourced 
human classification tasks. This choice stems from with our classification perspective 
(see \S\ref{ssec:classify_perspective}), and has been used in the prototype selection 
literature~\citep{bien2011prototype,kim2016examples}. Intuitively, 
a good set of prototype articles should uniquely identify a new article's group.

\noindent{\bf Datasets and features.}
{We {empirically} validate classification and prototype selection {methods on a well known USPS dataset}~\citep{bien2011prototype,kim2016examples}. 
USPS contains $16 \times 16$ grayscale handwritten digits in 10 classes (i.e., digits 0 through 9). To reduce the dimensionality we use PCA, projecting the 256 dimensional image vectors into 39 features that explain 85\% of the variance. 
The USPS dataset provides 7291 training and 2007 test images. We generate another 9 random splits with exactly the same number of training and test images for the purposes of estimating confidence intervals.}

{In using the USPS dataset our aims are twofold. First, it shows the versatility of the method: the domain need not be text, collections need not be separated by time, and it operates with more than two classes. Indeed, by thinking of each digit as a \textit{group}, our method can identify representative and diverse examples of digits. Second, our method can be seen as a special kind of prototype selection for which the USPS dataset has been used as a standard benchmark ~\citep{bien2011prototype,kim2016examples}.}

We further use the controversial news dataset described in \S\ref{sec:datasets} to evaluate comparative summarisation. 
We adopt the pre-trained GloVe-300 \citep{pennington2014glove} vector representation for each word, and then represent the article as an average of the word vectors from its the title and first 3 sentences 
-- the most important text due to the inverted pyramid structure in news style ~\citep{po2003news}. 
This feature performs competitively in retrieval tasks despite its simplicity~\citep{joulin2016bag}. 
For each news topic, we generate 10 random splits with 80\% training articles and 20\% test articles for automatic evaluation. One of these splits is used for human evaluation. 


\noindent{\bf Approaches and baselines.} We compare:
\begin{itemize}
\item {\it nn-comp-greedy} represents the nearest neighbour objective $\UCal_{\text{\em nn}}$, optimised in a greedy manner.  
\item {\it mmd-diff} represents the difference of MMD objective $\UCal_{\text{\em diff}}$. {\it mmd-diff-grad} uses gradient based optimisation while {\it mmd-diff-greedy} is optimised greedily. 
\item {\it mmd-div-grad} and {\it mmd-div-greedy} are the gradient-based and greedy variants of the diverse MMD objective $\UCal_{\text{\em div}}$.
\end{itemize}
with three baseline approaches:
\begin{itemize}
\item {\it kmeans} clusters with kmeans++ initialisation~\citep{arthur2007k} 
found separately for each document group. 
The $M$ cluster centers for each group are snapped to the nearest data point using Equation~\ref{equ:snap_point}.
\item {\it kmedoids}~\citep{kaufman1987clustering} clustering algorithm with kmeans++ initialisation, computed separately for each document group. The medoids become the prototypes themselves.
\item {\it mmd-critic}~\citep{kim2016examples} selects prototypes using greedy optimisation of $\mathrm{MMD}^2$ and criticisms by choosing points that deviate from the prototypes. The summary is selected from the unlabeled training set and consists of prototypes and criticisms in a one-to-one ratio.
\end{itemize}

We use the Radial Basis Function (RBF) kernel when applicable. The hyper-parameter $\gamma$ is chosen along with the trade-off factor $\lambda$, and SVM soft margin $C$ using grid search 3 fold cross-validation on the training set. 
Note that 1NN has no tunable parameters.
The {\it grad} optimisation approach uses the L-BFGS algorithm~\citep{byrd1995limited}, with initial prototype guesses chosen by the {\it greedy} algorithm for news dataset and K-means for USPS dataset.

\subsection{Automatic Evaluation Settings}

The controversial news dataset topics are divided into two groups 
of equal duration based on article timestamp. 
Note that typically the number of documents in each time range is imbalanced. 
The USPS hand written digits dataset is divided into 10 groups corresponding to the 10 different digits.
On each training split we select the prototypes for each group
and then train an SVM or 1NN on the set of prototypes. 

We measure the classifier performance on the test set using balanced accuracy,
defined as the average accuracy of all classes \citep{brodersen2010balanced}. 
For binary classification this is $\frac{1}{2}(\frac{\mathrm{TP}}{\mathrm{P}} + \frac{\mathrm{TN}}{\mathrm{N}} )$, 
{defined in terms of total positives $\mathrm{P}$, total negatives $\mathrm{N}$, true negatives $\mathrm{TN}$, and true positives $\mathrm{TP}$.}
Balanced accuracy accounts for class imbalance, and is applicable to both binary and multi-class classification tasks (whereas AUC and average precision are not). 
For all approaches, we report the mean and 95\% confidence interval of the 10 random splits. 

We report results on 2, 4, 8, or 16 prototypes per group -- a small number of prototypes is necessary for the summaries to be meaningful to humans. 
This is in contrast to the hundreds of prototypes used by ~\cite{bien2011prototype,kim2016examples}, in automatic evaluations of
the predictive quality of prototypes.

\subsection{Automatic Evaluation Results}\label{ssec::auto_eval}








Figure~\ref{fig:auto_bar} reports balanced accuracy for all methods using SVM and 1-NN across different datasets and numbers of prototypes. 
On the USPS dataset, most methods perform well.
The differences are small, if at all distinguishable. 
{\it mmd-critic} performs poorly on USPS; 
this is because it does not guarantee 
a fixed number of prototypes per group, and sometimes misses a group all together. 
Note that this is very unlikely to occur with only 2 groups
in the news dataset.


On the three news datasets, comparative summaries based on \redo{ {\it mmd} objectives 
are the best-performing approach in 20 out of 24 evaluations} (2 classifiers x 4 prototype sizes x 3 news topics (details in Appendix$^1$)
. In the remaining two cases, they are the second-best with overlapping confidence intervals against the best ({\it kmeans}). 
Despite the lack of optimisation guarantees, {\it grad} optimisation produces 
prototypes of better quality in \redo{14} out of 24 settings. 

Generally, all methods produce better classification accuracy as the number of prototypes increases. This indicates that the chosen prototypes do introduce new information that helps with the classification. In the limit, where all documents are selected as prototypes -- a setting that is clearly unreasonable when summarisation is the goal -- the performance is determined by the classifier alone. SVM achieves 0.763 on \emph{Capital Punishment} and \emph{Beef Ban}, 0.707 on \emph{Gun Control}, while 1-NN achieves 0.762 on \emph{Capital Punishment}, 0.763 on \emph{Beef Ban} and 0.702 on \emph{Gun Control}. As seen in Figure~\ref{fig:auto_bar} no prototype selection method approaches this accuracy. 
This highlights the difficulty of selecting only a few prototypes to represent complex distributions of news articles over time.

\subsection{Crowd-sourced Evaluation Settings}
\label{ssec:fig8_setting}

{We conduct a user study on the crowd-sourcing platform figure-eight\footnote{https://www.figure-eight.com} with two questions in mind: 
(1) using article classification accuracy as a proxy, 
    do people perform similarly to automatic evaluation?
(2) how useful do people find the comparative summaries?}
{This is an acid test on providing value to users who need comprehend 
large document corpora. Human evaluations in this work are designed to grade 
our method in a real world task: accurately identifying a news articles group 
(e.g. the month it is published) given only a few (4) articles from each month. 
The automatic evaluations in 
\S\ref{ssec::auto_eval} are instructive proxies for efficacy, but 
inherently incomplete without human evaluation.
}

\noindent{\bf Generating summaries for the crowd.} We present summaries from four methods {\it kmeans}, {\it nn-comp-greedy}, {\it mmd-diff-greedy}, and {\it mmd-diff-grad} -- chosen because they perform well in automatic evaluation and together form a cross-section of different method types.
We opt to vary the groups of news articles being summarised by choosing many pairs of time ranges,
since summaries on the same pair of groups (by definition) 
tend to be very similar or identical, which incurs user fatigue. 
{We use the \textit{Beef Ban} topic because it has the longest time range: 
June 2017 to July 2018 inclusive. 
The articles are grouped into each of the 14 months, 
and then 91 (i.e., 14 choose 2) pairs are formed. }
We take the top 10 pairs by performance according automatic evaluation 
using each of the four approaches, the union of these lead to 21 pairs.
We pick top-performing pairs because preliminary human experiments showed 
that humans seem unable to classify an article when automatic results do poorly 
(e.g. $<$0.65 in balanced accuracy).
Articles from each of the 21 pairs of months are randomly split into training and testing sets. We ask participants to classify six randomly sampled test articles. 
To reduce evaluation variance, all methods share the same test articles, 
different methods are randomized and are blind to workers. 
We record three independent judgments for each (test article, month-pair) tuple 
-- totaling 1,512 judgments from 126 test questions over four methods. 
We also restrict the crowd workers to be from India, 
where \textit{Beef Ban} is locally relevant,  
and workers will be familiar with the people, places and organisations mentioned news articles. 

\begin{figure}[tb]
    \centering
    \includegraphics[width=0.96\linewidth]{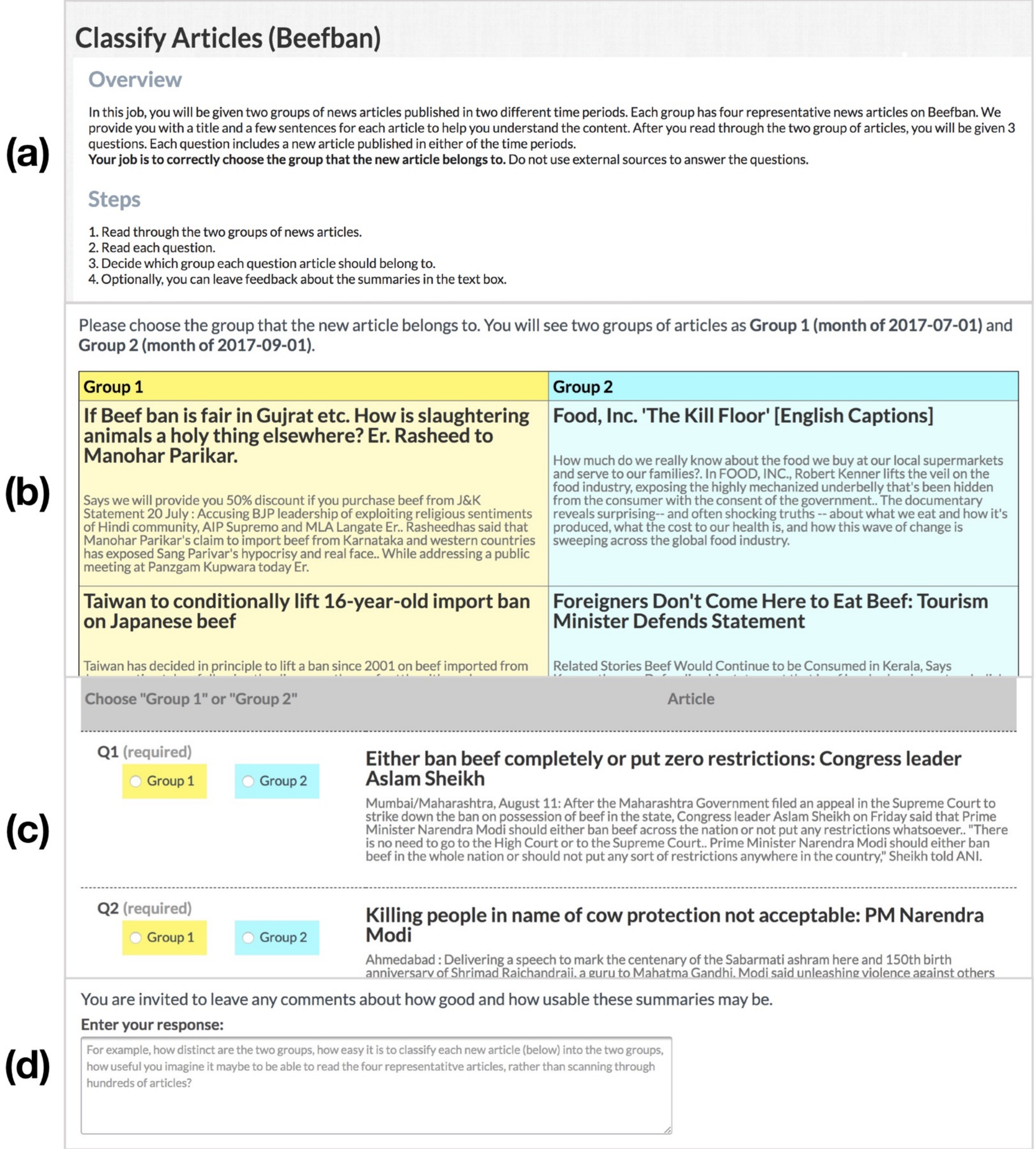}
    \caption{An example questionnaire used for crowd-sourced evaluation. It consists of: (a) instructions, (b) two groups of summaries,
    (c) question articles, and (d) a comment box for feedback.
    See \S\ref{ssec:fig8_setting}. 
    }
    \label{fig:fig8}
\end{figure}

\noindent\textbf{Questionnaire design.} 
Figure~\ref{fig:fig8} shows the questionnaires we designed for human evaluation. Each questionnaire has 4 parts: (a) instructions, (b) two groups of prototypes, (c) test articles that must be classified into a group, and (c) a comment box for free-form feedback.

In the instruction (a), we explain that the two groups of representative articles (the prototypes for each time range) are articles from different time ranges and lay out the steps to complete the questionnaire. We ask participants not to use external sources to help classify test articles. 

The two groups of prototype articles (b) are chosen by one of the method being evaluated (e.g., {\it mmd-diff-grad} or {\it kmeans}) from articles in two different time ranges. Each group has four representative articles and each article has a title and a couple of sentences to help understand the content. We assign a different background colour to each group of summaries to give participants a visual guide.

Below the groups of summary articles are three questions (c), though for brevity only two are shown in Figure~\ref{fig:fig8}. Each question asks participants to decide which of the two time ranges a test article belongs to. 

We add a comment box (d) to gather free-form feedback from participants. This helps to quickly uncover problems with the task, provides valuable insight into how participants use the summaries to make their choices, and gives an indication of how difficult users find the task.
As a quality-control measure, 
we include questions with known ground truth amongst the test questions. These ground truth questions are manually curated and reviewed if many workers fail on them. 
Each unit of work includes 4 questionnaires (of 3 questions each), one of which is a group of ground truth questions randomly positioned. 
Note that ground truth questions are only used to filter out participants 
and are not included in the evaluation results. 

\subsection{Crowd-sourced Evaluation Results}
\label{ssec:fig8_results}

\noindent{\bf Worker profile.} The number of unique participants answering test questions ranged from 25 to 31 for each method,  
indicating that the results were not dominated a small number of participants. 
On average, participants spent 51 seconds on each test question and 2 minutes 33 seconds on each summary.

\begin{figure}[tb]
    \centering
     \includegraphics[height=3.7cm]{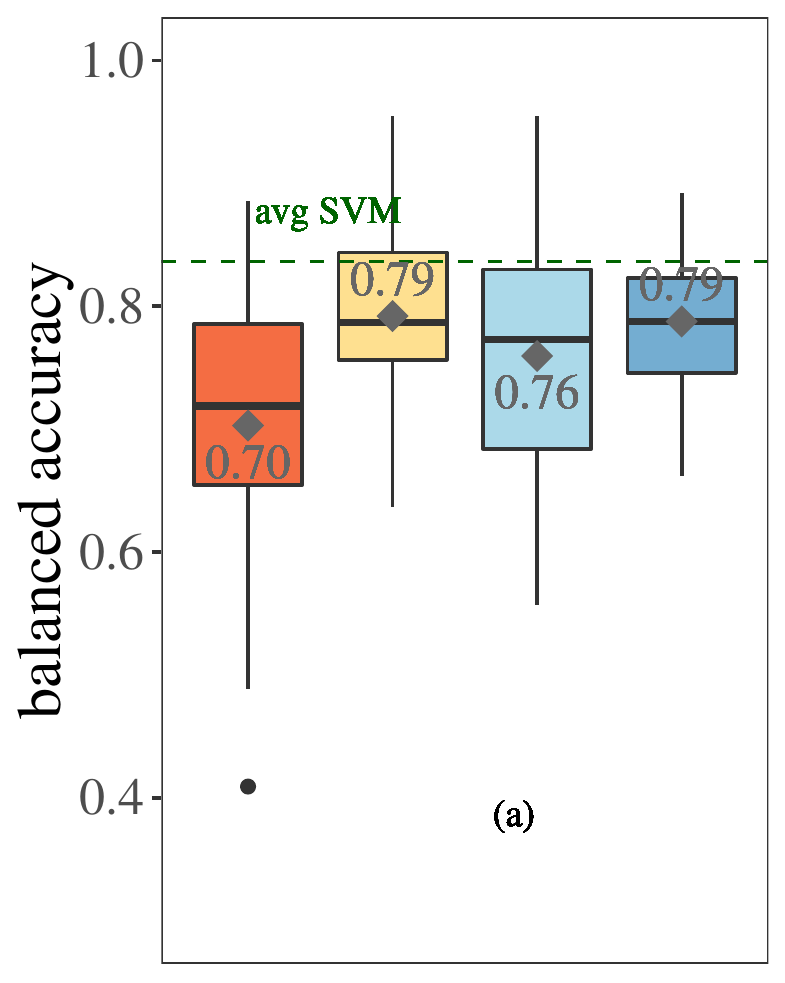}
     \includegraphics[height=3.7cm]{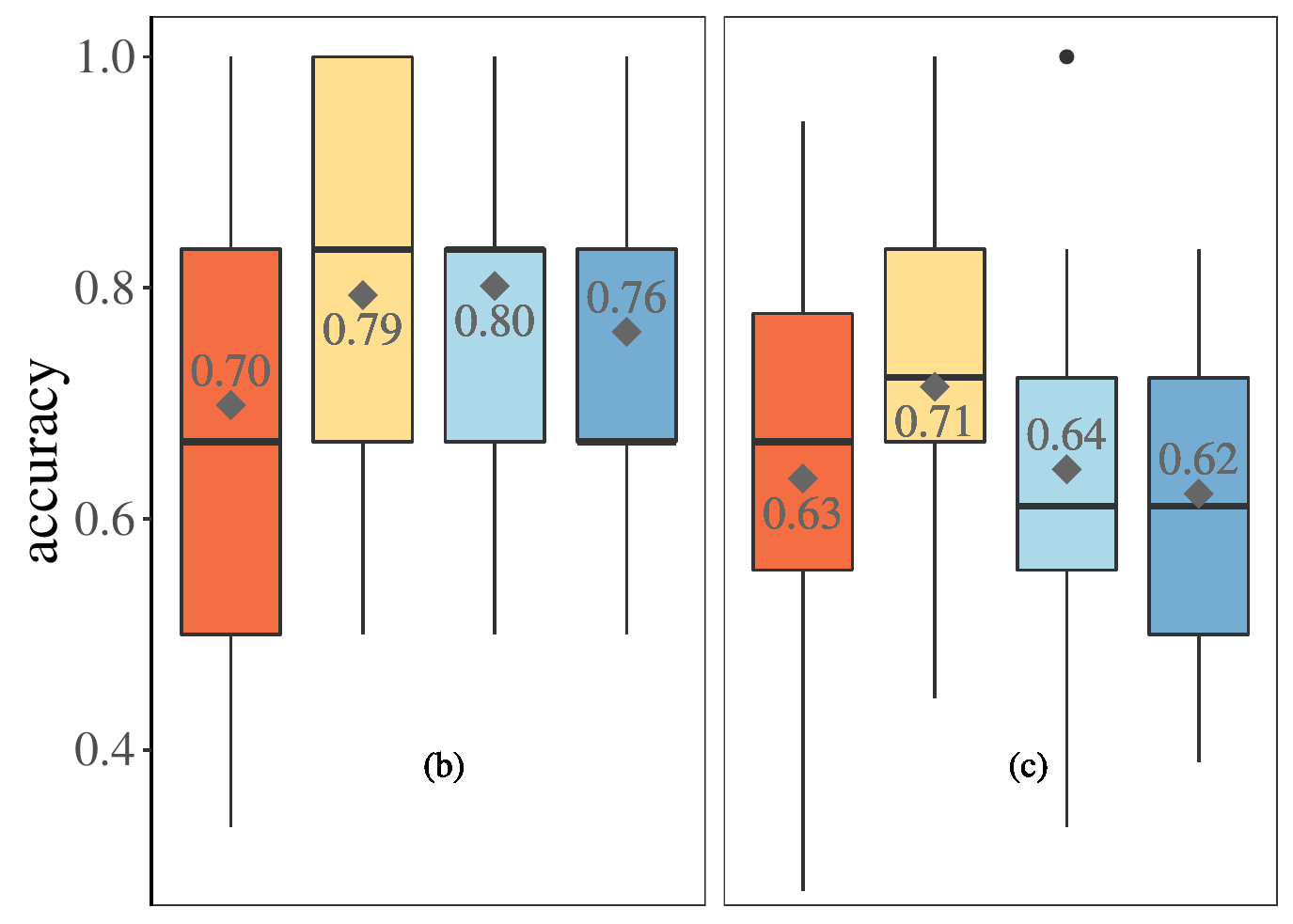}
     \includegraphics[width=0.98\linewidth]{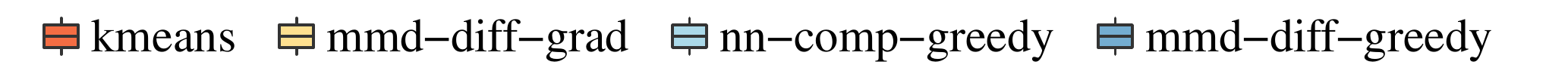}
    \caption{Classification accuracies for 21 pairs of summaries. (a) Automatic classification using prototypes (by SVM) on the entire test set. The green \textit{avg SVM} line is the mean accuracy of SVMs trained on the entire training set. (b) Automatic classification evaluated on 6 test articles per pair. (c) Human classification accuracy on 6 test articles per pair.}
    \label{fig:fig8_box}
\end{figure}
\noindent\textbf{Quantitative results}
Figure~\ref{fig:fig8_box} shows that on average crowd workers with {\it mmd-diff-grad} summaries classify 
an article more accurately than summaries from other approaches by at least 7\%.
{The results are statistically significant with $p<0.05$ under a one sided 
\emph{sign test}; which applies because the 126 test questions where answered 
by three random people and we cannot assume normality.}
It also has the highest number of consensus correct judgments (details in Appendix$^1$).
{\it mmd-diff-greedy} performs worse than {\it mmd-diff-grad}. 

We also compute the Fleiss Kappa statistic to measure inter-annotator agreement. 
The statistics are: 0.418 for {\it kmeans}, 0.456 for {\it mmd-diff-grad}, 
0.435 for {\it nn-comp-greedy}, and 0.483 for {\it mmd-diff-greedy} 
and a combined statistic of 0.451. All statistics fall into the range of 
moderate agreement ~\citep{landis1977measurement}, 
{which means the results we obtain in crowdsourced evaluations are reliable}.

The good performance of gradient-based optimisation is surprising given greedy approaches are usually preferred in subset selection tasks, due to approximation guarantees for submodular objectives. 
One plausible explanation is 
that early prototypes selected by {\it greedy} tend to cluster around the first prototype, whereas the simultaneous optimisation in {\it grad} tend to spread prototypes in feature space. With only four prototypes being shown to users, diversity is an important factor for human classification. 
Previous studies of {\it greedy} methods for prototype selection have used hundreds of 
prototypes~\citep{bien2011prototype} -- a setting in which the diversity of the early prototypes matters less -- or used 
criticisms~\citep{kim2016examples} to improve diversity in tandem.

Comparing Figure~\ref{fig:fig8_box} (a) -- (c), automatic classifiers trained on both the entire training set and prototypes have higher classification accuracy than human workers across all methods. This observation indicates that using summaries to classify articles is difficult for humans. It could also indicate that humans use different features for article grouping, and word vectors alone may not capture those features. 


\noindent{\bf Qualitative observations.}
Results from the optional free-form comments show that the participants found the classification difficulty to vary wildly. While some sets of articles were apparently easy to classify (e.g., ``Group articles are distinct in their manner, among which all are articles are easy to determine."), other articles were difficult to classify (e.g., ``Although two groups are clearly distinct, this one (news article) was pretty difficult to ascertain in which group it belongs to.") In some cases poor summaries seem to have made the task exceedingly difficult; e.g., ``Q1, Q2, Q3 all are not belongs to group 1 and group 2 any topic I think." (quoted verbatim).

We found that the {\it Beef Ban} topic interested many of our participants, with some expressing their views on the summarised articles, for example ``Firstly we should define what is beef ..is it a cow or any animal?" and ``It is a broad matter, what we should eat or not, it cannot be decided by government." (edited for clarity).

Participant comments also give some insight into what features were used to make classification. In particular, word and entity matching were frequently mentioned, a representative user comment is ``None of the questions match the given article, but I had to go by words used." All crowd-sourced evaluation results and comments are available in the dataset github repository$^1$.







\section{Conclusion}

We formulated the comparative document summarisation in terms of competing binary classifiers. 
This inspired new $\mathrm{MMD}$ based objectives 
amenable to both gradient and greedy optimisation. 
Moreover, the setting enabled us to design efficient automatic and human evaluations 
to compare different objectives and optimisation methods on a new, highly relevant dataset of news articles. 
We found that our new $\mathrm{MMD}$ approaches, optimised by gradient methods, frequently outperformed all alternatives, including the greedy approaches currently favoured by the literature. 
Future work can include new use cases for comparative summarisation, such as authors or view points; richer text features; extensions to cross-modal comparative summarisation. 


\paragraph{\bf Acknowledgements.}
This work is supported by the ARC Discovery Project DP180101985. 
This research is also supported by use of the NeCTAR Research Cloud, 
a collaborative Australian research platform supported by the National Collaborative Research Infrastructure Strategy.

{
\fontsize{9.0pt}{10.0pt}
  \bibliographystyle{aaai}
\bibliography{main.bib}
}
\onecolumn
\appendix 
\part*{Supplementary Appendix}
This supplementary information contains additional information required for the main paper, including details of greedy and gradient optimisation, 
analysis of crowdsourced evaluations and results table of automatic evaluations. The source code and datasets of this work is available online. \footnote{https://github.com/computationalmedia/compsumm}
\section{Greedy Algorithm}\label{suppl:greedy}
The greedy approach to maximise a utility function $\UCal(\bar{\mathbf{X}})$ is outlined in Algorithm 1. 
We iterate through all groups, adding one point to the summary at a time, selected as one that brings the largest marginal gain to the utility function $\UCal(\cdot)$. 
Details necessary for computing the marginal gain in step 7 are found in \S~\ref{suppl:deriv}. 

\begin{algorithm}
 \label{algo:greedy}
    \caption{Greedy algorithm to maximize
    objective $\UCal(\cdot)$}
    \label{alg:mmd_greedy}
    \begin{algorithmic}[1]
        \Require{$\{ \mathbf{X}_{1:G} \}$: Groups of documents, \\
        $G$: number of groups, $M$: number of prototypes per group}
        \Procedure{GreedyMax}{$\mathbf{X}, G, M$}
        \State $(\forall {g \in 1 \ldots G})  \, \bar{\mathbf{X}}_g \gets \{ \}$
        
        \For{m from 1 to M}
        	\State $\bar{\mathbf{X}} \gets \cup_{g=1}^{G} \mathbf{\bar X}_g$
        	\For{g from 1 to G}
          	\State $\bar{\x}_{g,m} \gets
              \underset{\x_g  \in \mathbf{X}_g \setminus \bar{\mathbf{X}}_g}{\mathrm{argmax}} \Delta_\UCal(\x_g | \bar{\mathbf{X}})$
              \State $\bar{\mathbf{X}}_g \gets \bar{\mathbf{X}}_g \cup \bar\x_{g,m}$ 
            \EndFor
        \EndFor 
        \State $\bar{\mathbf{X}} \gets \cup_{g=1}^{G} \mathbf{X}_g$
        	\State \textbf{return} $\bar{\mathbf{\bar X}}$ 
         \EndProcedure
    \end{algorithmic}
\end{algorithm}

\section{Gradients and Discrete Derivatives of the MMD objectives}\label{suppl:deriv}
The equation and gradient gradient of $MMD^2(\bar{\mathbf{A}}_g, \mathbf{X}_g)$ for the RBF Kernel are defined as:
\begin{align}
    MMD^2(\bar{\mathbf{A}}_g, \mathbf{X}_g) =& -\frac{2}{M \times N_g}\sum_{i=1}^{N_g}\sum_{j=1}^{M_g} k( \bar{\pmb a}_{g,j}, \x_{g,i}) + \frac{1}{M_g^2}\sum_{i, j = 1}^{M_g} k(\bar{\pmb a}_{g,i}, \bar{\pmb a}_{g,j}) \label{eqn::mmd_obj}\\
    \forall_{l \in 1\ldots M_g} \nabla_{\bar{\pmb a}_{g,l}} MMD^2(\bar{\mathbf{A}}_g, \mathbf{X}_g) =& \frac{4\gamma}{M_g} \bigg(-\frac{1}{N_g} \sum_{i = 1}^{N_g} {k(\bar{\pmb a}_{g,l}, \x_i)(\x_i - \bar{\pmb a}_{g,l}) } + \frac{1}{M_g^2} \sum_{i = 1}^{M_g} {k(\bar{\pmb a}_{g,i}, \bar{\pmb a}_{g,l})(\bar{\pmb a}_{g,i} - \bar{\pmb a}_{g,l}) }  \bigg) \label{eqn::mmd_grad}
\end{align}

$MMD^2(\mathbf{A}_g, \mathbf{X}_{\lnot g})$ can also be computed in a similar way, by replacing $\x_{g,i}$ by $\x_{\lnot g, i}$ in equation \eqref{eqn::mmd_grad}, this will yield the objective of $\UCal_{\text{\em diff}}(\bar{\mathbf{X}})$ \eqref{equ:mmd_labels}. The first term of the equation \eqref{eqn::mmd_grad} corresponds to the gradient of first term of equation \eqref{eqn::mmd_obj}. Hence, it will yield the gradient of the objective $\UCal_{\text{\em div}}(\bar{\mathbf{X}})$ \eqref{equ:mmd_labels_div}.

Let $V_g$ be the indices of $\mathbf{X}_g$ and $S_g$ be the indices of $\bar{\mathbf{X}}_g$. The discrete derivatives for $-MMD^2(\bar{X}_g, X_g)$ is.

\begin{align}
	\Delta_{-MMD^2(\bar{\mathbf{X}}_g, \mathbf{X}_g)}(\x_g \mid \bar{\mathbf{X}}_g) =& \frac{1}{|S_g| + 1}  \bigg( \frac{2}{|V_g|} \sum_{i \in V_g} k(\x_i, \x_g) - \frac{2}{|V_g| |S_g|} \sum_{i \in V_g, j \in S_g} k(\x_i, \x_j) + \nonumber \\& \frac{2|S_g|+1}{|S_g|^2(|S_g| + 1)}\sum_{i, j \in S_g}k(\x_i, \x_j) - \frac{2}{(|S_g| + 1)}\sum_{i \in S_g} k(\x_i, \x_g) - \frac{1}{|S_g| + 1}k(\x_g, \x_g)  \bigg) \label{equ:mmd_disc_deriv}
\end{align}
Discrete derivatives of different MMD objectives (equations \ref{equ:mmd_labels}, \ref{equ:mmd_labels_div}) can be built upon equation \eqref{equ:mmd_disc_deriv}. The discrete derivative of $\lambda$ term in equation \eqref{equ:mmd_labels_div} is given by first two terms of equation \eqref{equ:mmd_disc_deriv}. The Discrete derivative of equation \eqref{eq:nncomp} can be computed in a similar way. Equations for discrete derivatives allow greedy optimisation (\S \ref{suppl:greedy}) to be done efficiently.




\section{Methodology for data collection}\label{suppl:dataset}

\noindent\textbf{Topic curation.}
We curate an initial list of 10 topics in June 2017 that satisfying
vthe criteria of having non-trivial news coverage and being controversial. In this work, we use \textit{Beef Ban}, \textit{Capital Punishment} and \textit{Gun Control} topics. The other topics are \textit{Climate change, Illegal immigration, Refugees, Gay marriage, Animal testing, Cyclists on road} and \textit{Marijuana}.
We want to focus on controversial topics since they are likely to be discussed in the future since their coverage lasts for a long time.
Controversy is an important topic for research in social media and
online political discourse, is also important in real-world applications such as intelligence and business strategy development.

To obtain various opinions on contemporary social problems, we choose
Twitter as a source since it is frequently used for reporting and sharing related
news articles. \cite{garimella2018quantifying} use similar approach
generating Twitter dataset on the controversial topics. The authors consider
Twitter hashtags as query and use similarity function to retrieve similar hashtags.
We obtain embedded news articles from Twitter posts to generate a dataset
and use a different expansion approach to retrieve related hashtags.

\begin{table}[!htbp]
  \centering\small
   \begin{tabular}{| l | l | r | r | r |}
   \hline
   Topic & Queries & \#Tweets & \#News & \#News (cleaned) \\
   \hline
   \emph{Beef Ban} & beef ban, beefban & 304,234 & 17,131 & 1,543 \\
   \emph{Capital Punishment} & death penalty, deathpenalty, capital punishment & 11,052,295 & 66,542 & 7,905 \\
   \emph{Gun Control} & gun control, guncontrol, gunsense, gunsafety, & 36,533,525 & 130,312 & 6,494 \\
   &  gun laws, gun violence & & &\\
   \hline
   \end{tabular}
 \caption{Controversial Topic Dataset Statistics}
 \label{table::newsdataset}
\end{table}

\noindent\textbf{Query curation.}
We use a hashtags expansion approach \citep{verkamp2013five} to curate
relevant queries for each topic. We first manually select a single query for each topic,
then use it to collect Twitter posts for two weeks. These posts are used as
an initial data set that we create a query set based on. We extract the 10 most
common hashtags that appear in the initial dataset. These hashtags are used
to query the same dataset again and then we re-extract the 10 most common
hashtags from the query result. We continue this iteration several times
until the hashtags used for query and the re-extracted hashtags are the same.
All of the topics finish generating a query set after 4\texttt{\char`\~}5 iterations.

Based on the query set generated using the hashtags expansion, we perform
additional filtering. Location hashtags such as \#Florida or \#Alabama are removed
to prevent detailed locations being discussed.
Some hashtags like \#cow, \#beef, \#PJNET, and \#2A are excluded
since they are not directly related to the topics or are too general.
As a result, \textit{Beef Ban} topic is defined by a single query while
\textit{Capital Punishment} and \textit{Gun Control} include more diverse hashtags in
the query set. Table \ref{table::newsdataset} summaries the query set used for each topic.

\noindent\textbf{Article extraction.}
After generating a query set for each topic, we fetch the Twitter stream that includes
any of the hashtags in the query set. Twitter post frequently includes embedded
news articles related to the post. We focus on the news articles in this work
since they generally include more coherent stories than the corresponding
Twitter post dataset. We extract the embedded news articles
by visiting the article URL and downloading the content from it.
By doing that, we can collect news articles that are mentioned and shared
in ongoing social media which can be a measure of how important and accurate the news is.
We clean the data by filtering spam articles and removing duplicate articles mentioned in multiple
tweets. To increase the relevance, we remove garbage texts such as
"Subscribe to our channel", "Please sign up" or "All Rights Reserved"
that repeatedly appear with the news content. Table \ref{table::newsdataset} reports the number of the Tweets and the news articles before and after the cleaning.

\section{Human Evaluation Results}\label{suppl:results:human}

\begin{table*}
\centering
\begin{tabular}{r|rrr}
    		& \#unique & correct by & correct  \\
Method     	& workers & majority & judgements  \\
\hline 
{\it kmeans}     & 31             & 81                                 & 240                 \\
{\it mmd-diff-grad}   & 25             & 94                                 & 270                 \\
{\it nn-comp-greedy}    & 28             & 80                                 & 243                 \\
{\it mmd-diff-greedy} & 29             & 83                                 & 235                 \\
\hline
Total      & 40             & 126                                & 378                
\end{tabular}
\caption{Results of Human Pilot Study on Classification Task. Unique workers participating in classifying test articles for each method is in first column. Correct by majority means the number of test articles (out of 126) classified correctly by majority (at least two people). Correct Judgments indicates the number of individual judgments that are correct (out of 378)}
\label{tab:human_eval_results}
\end{table*}

\begin{figure}[h]
    \centering
    \includegraphics[width=0.5\linewidth]{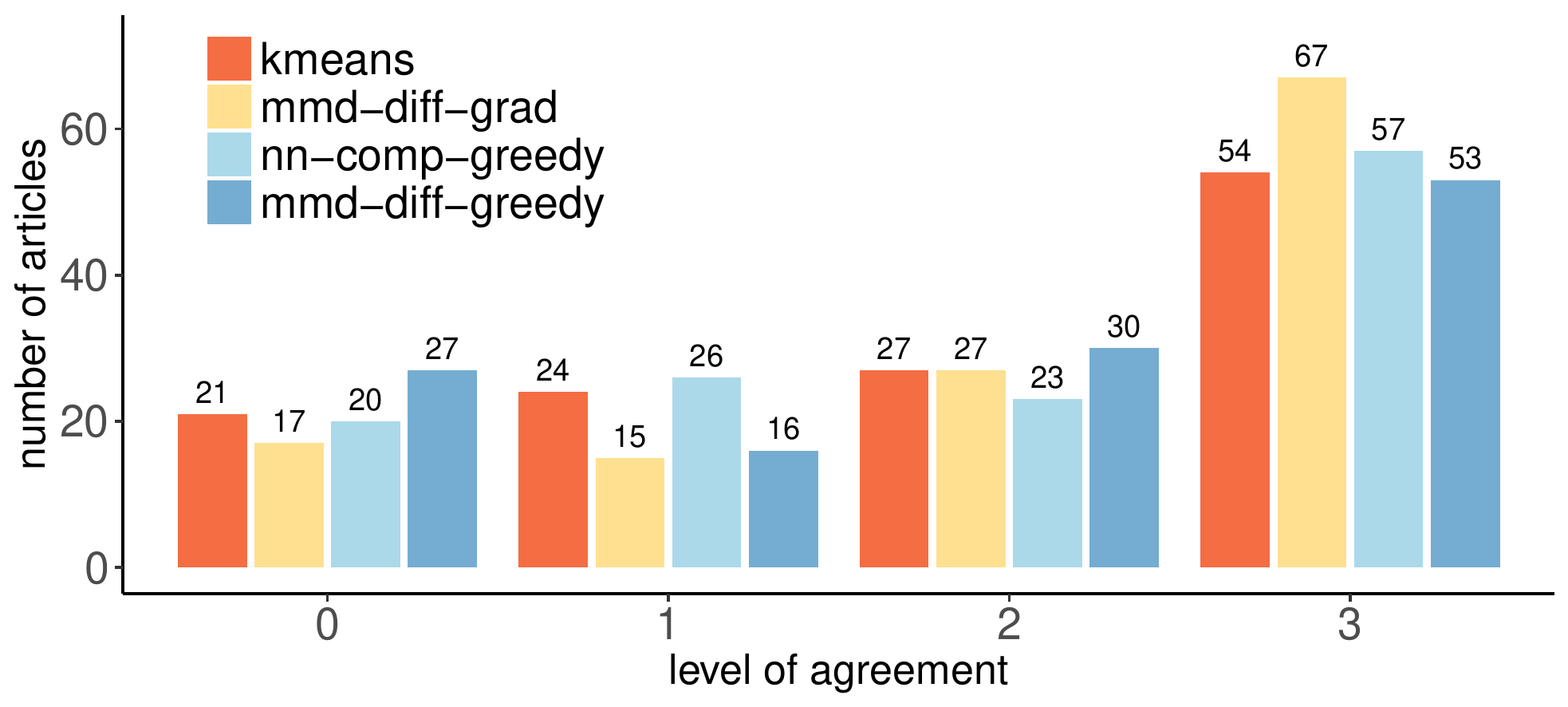}
    \caption{Shows the number of test articles that humans correctly classified, at different agreement levels. A level of 3 means all participants correctly classified the test article, while 0 means all participants incorrectly classified the test article.}
    \label{fig:fig8_perf}
\end{figure}

\begin{figure}
    \centering
    \includegraphics[width=0.6\linewidth]{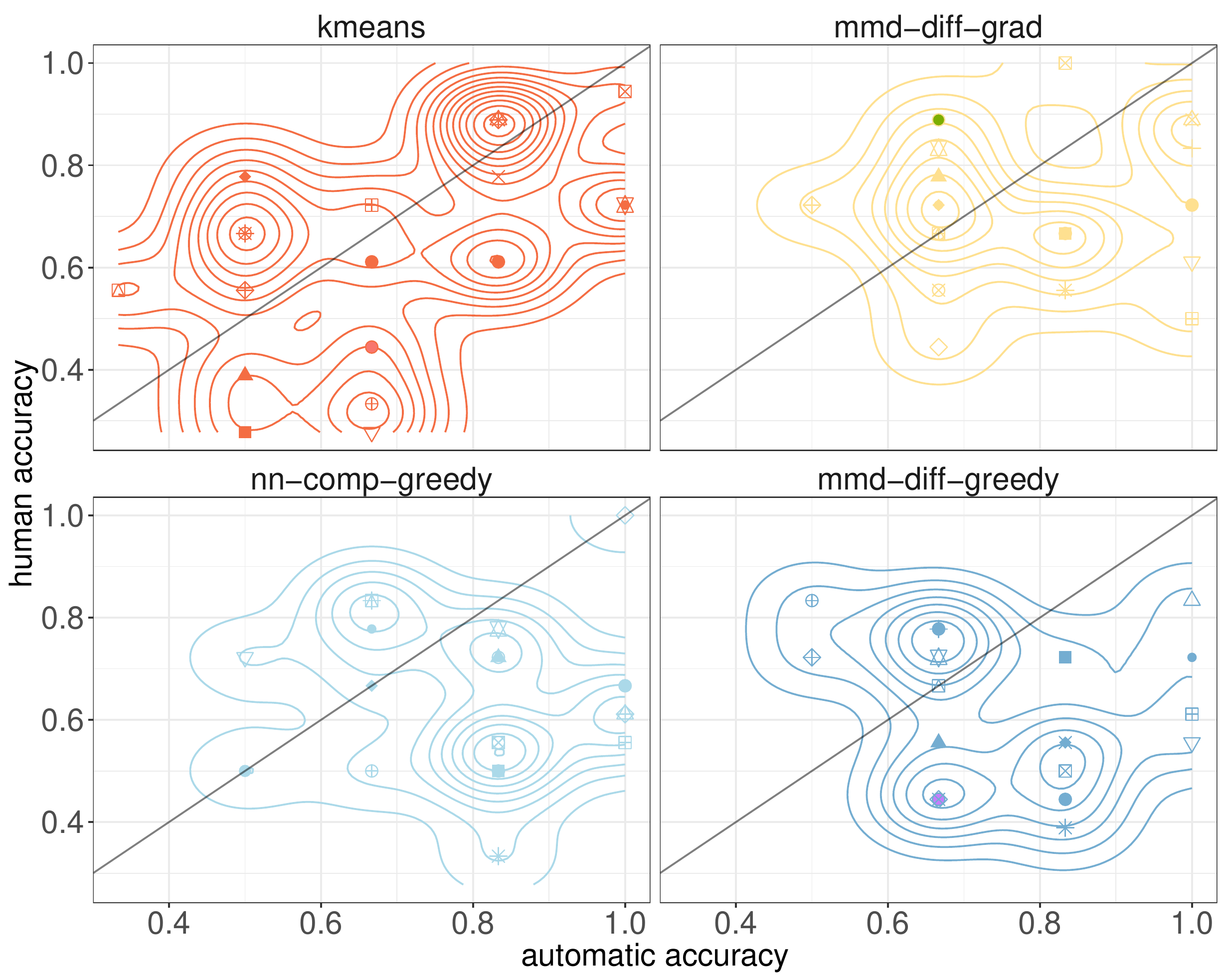}
    \caption{Density plot of human accuracy vs. machine classifier (SVM) accuracy. Each point means a pair of summaries and we average the accuracy over the 6 test articles.
    People are doing roughly same as SVM for {\it kmeans} and {\it mmd-diff-grad}, whereas people seems to do worse than SVM in case of greedy methods.}
    \label{fig:fig8_density}
\end{figure}

\textit{This version (version 2) fixes the bug in SVM automatic evaluation. 
In version 1, hyperparameter $C$, was not being chosen due to a bug in the source code. Overall, there is only few minor changes in the results. 
The updated automatic evaluation table is provided in \S~\ref{suppl:results_auto}.}

Table \ref{tab:human_eval_results} shows the number of unique participants and the number of correct judgements for each methods from the human evaluation result. 
The number of unique participants answering test questions ranged from 25 to 31. The union of the participants involved in any four methods is 40.
The number of articles correctly classified by the participants when we evaluate by majority voting and treat each judgment shows the efficacy of the proposed method {\it mmd-diff-grad} over other methods including {\it kmeans} baseline. 

Figure~\ref{fig:fig8_perf} shows the level of agreement across participants for each method. First we note that participants were frequently able to complete the task of classifying new articles correctly into one of two groups, this is shown by the large fraction of articles for which the correct group was unanimously chosen. Compared to other comparative prototype selection methods {\it mmd-diff-grad} has the largest number of articles correctly classified by all three participants, beating the next best {\it nn-comp-greedy} by 10 articles. Consequently {\it mmd-diff-grad} also has fewer articles which were unanimously assigned to the incorrect group by participants.

Figure \ref{fig:fig8_density} shows the density plot comparing human accuracy vs. SVM accuracy in classifying small set of test articles we had for human evaluation. 

\section{Results Table}\label{suppl:results_auto}

Automatic evaluation results table corresponding to error-bar plot in Figure \ref{fig:auto_bar} is given in Tables \ref{auto_capital}, \ref{auto_beefban}, \ref{auto_guncontrol} and \ref{auto_usps}, 
We report the mean of balanced accuracy and 95\% confidence interval of the 10 random splits. The left four columns use 1-NN and the right four use SVM. We report results on 2, 4, 8, or 16 prototypes per group. The highest performing method is in bold and second highest performing one is in italics.


\begin{table}
\resizebox{\linewidth}{!}{%
\centering\footnotesize
\begin{tabular}{r|llll|llll}
method & 2 & 4 & 8 & 16 & 2 & 4 & 8 & 16 \\
\hline
kmedoids 
& 0.501 $\pm$ .010  & 0.505 $\pm$ .011  & 0.528 $\pm$ .012  & 0.530 $\pm$ .021  
& 0.498 $\pm$ .008  & 0.501 $\pm$ .005  & 0.523 $\pm$ .014  & 0.523 $\pm$ .02  \\ 
kmeans 
& 0.510 $\pm$ .017  & 0.542 $\pm$ .012  & \em 0.546 $\pm$ .007  & 0.576 $\pm$ .008  
& 0.507 $\pm$ .016  & \em 0.546 $\pm$ .009  & 0.541 $\pm$ .01  & 0.551 $\pm$ .012  \\ 
mmd-critic
& 0.499 $\pm$ .003  & 0.514 $\pm$ .016  & 0.515 $\pm$ .013  & 0.531 $\pm$ .009  
& 0.5 $\pm$ .001  & 0.499 $\pm$ .01  & 0.51 $\pm$ .017  & 0.53 $\pm$ .012  \\ 
mmd-diff-grad
& \em 0.534 $\pm$ .011  & \em 0.538 $\pm$ .014  & 0.540 $\pm$ .020  & \bf 0.582 $\pm$ .010 
& \em 0.53 $\pm$ .011  & 0.542 $\pm$ .01  & 0.559 $\pm$ .012  & 0.575 $\pm$ .011  \\ 
mmd-div-grad
& \bf 0.539 $\pm$ .013  & \bf 0.545 $\pm$ .008  & \bf 0.564 $\pm$ .014  & \em 0.579 $\pm$ .011  
& 0.527 $\pm$ .012  & 0.546 $\pm$ .014  & \em 0.56 $\pm$ .012  & \bf 0.579 $\pm$ .009  \\ 
nn-comp-greedy
& 0.509 $\pm$ .011  & 0.515 $\pm$ .008  & 0.544 $\pm$ .009  & 0.577 $\pm$ .007  
& 0.517 $\pm$ .018  & 0.529 $\pm$ .015  & \bf 0.56 $\pm$ .007  & 0.577 $\pm$ .006  \\ 
mmd-diff-greedy
& 0.530 $\pm$ .009  & 0.536 $\pm$ .012  & 0.545 $\pm$ .013  & 0.564 $\pm$ .013  
& 0.529 $\pm$ .014  & \bf 0.554 $\pm$ .012  & 0.553 $\pm$ .012  & 0.575 $\pm$ .011  \\ 
mmd-div-greedy
& 0.530 $\pm$ .010  & 0.525 $\pm$ .011  & 0.539 $\pm$ .012  & 0.563 $\pm$ .009 
& \bf 0.532 $\pm$ .014  & 0.544 $\pm$ .013  & 0.544 $\pm$ .007  & \em 0.578 $\pm$ .011 \\
\hline
\end{tabular}
}
\caption{Classification performance on {\it Capital Punishment} News dataset. (left) 1-NN, (right) SVM.}
\label{auto_capital}
\vspace{.5cm}
\resizebox{\linewidth}{!}{%
\centering\footnotesize
\begin{tabular}{r|llll|llll}
method & 2 & 4 & 8 & 16 & 2 & 4 & 8 & 16\\
\hline
kmedoids
& 0.592 $\pm$ .016  & 0.586 $\pm$ .011  & 0.582 $\pm$ .025  & 0.589 $\pm$ .023 
& 0.573 $\pm$ .022  & 0.571 $\pm$ .013  & 0.574 $\pm$ .025  & 0.578 $\pm$ .031  \\ 
kmeans 
& 0.582 $\pm$ .020  & \bf 0.613 $\pm$ .017  & \em 0.622 $\pm$ .030  & 0.629 $\pm$ .025  
& 0.57 $\pm$ .024  & \em 0.608 $\pm$ .022  & \em 0.615 $\pm$ .02  & 0.637 $\pm$ .027  \\ 
mmd-critic
& 0.541 $\pm$ .031  & 0.524 $\pm$ .028  & 0.530 $\pm$ .025  & 0.535 $\pm$ .033   
& 0.531 $\pm$ .028  & 0.533 $\pm$ .021  & 0.516 $\pm$ .039  & 0.542 $\pm$ .029  \\ 
mmd-diff-grad
& \em 0.595 $\pm$ .019  & 0.587 $\pm$ .023  & 0.610 $\pm$ .017  & \em 0.633 $\pm$ .020  
& 0.592 $\pm$ .036  & \bf 0.612 $\pm$ .022  & 0.615 $\pm$ .024  & \em 0.641 $\pm$ .019  \\ 
mmd-div-grad
& \bf 0.602 $\pm$ .021  & \em 0.605 $\pm$ .015  & 0.605 $\pm$ .028  & \bf 0.636 $\pm$ .028  
& \em 0.595 $\pm$ .019  & 0.597 $\pm$ .025  & 0.608 $\pm$ .026  & 0.636 $\pm$ .019  \\ 
nn-comp-greedy
& 0.587 $\pm$ .018  & 0.600 $\pm$ .027  & \bf 0.624 $\pm$ .029  & 0.627 $\pm$ .026  
& 0.588 $\pm$ .02  & 0.605 $\pm$ .014  & 0.615 $\pm$ .023  & 0.639 $\pm$ .026  \\ 
mmd-diff-greedy
& 0.592 $\pm$ .027  & 0.595 $\pm$ .021  & 0.615 $\pm$ .019  & 0.629 $\pm$ .021  
& \bf 0.596 $\pm$ .019  & 0.597 $\pm$ .022  & \bf 0.626 $\pm$ .021  & \bf 0.642 $\pm$ .021  \\ 
mmd-div-greedy
& 0.586 $\pm$ .014  & 0.578 $\pm$ .021  & 0.593 $\pm$ .022  & 0.616 $\pm$ .025  
& 0.594 $\pm$ .019  & 0.583 $\pm$ .026  & 0.597 $\pm$ .027  & 0.632 $\pm$ .024  \\ 
\hline
\end{tabular}
}
\caption{Classification performance on {\it Beefban} News dataset. (left) 1-NN, (right) SVM.}
\label{auto_beefban}
\vspace{.5cm}
\resizebox{\linewidth}{!}{%
\centering\footnotesize
\begin{tabular}{r|llll|llll}
method & 2 & 4 & 8 & 16 & 2 & 4 & 8 & 16\\
\hline
kmedoids
& 0.501 $\pm$ .016  & 0.504 $\pm$ .012  & 0.518 $\pm$ .016  & 0.518 $\pm$ .013 
& 0.502 $\pm$ .014  & 0.502 $\pm$ .007  & 0.516 $\pm$ .011  & 0.52 $\pm$ .006  \\ 
kmeans 
& 0.505 $\pm$ .009  & 0.506 $\pm$ .006  & \bf 0.538 $\pm$ .013  & 0.542 $\pm$ .014 
& 0.508 $\pm$ .007  & 0.504 $\pm$ .01  & 0.527 $\pm$ .01  & 0.536 $\pm$ .008  \\ 
mmd-critic 
& 0.506 $\pm$ .006  & 0.511 $\pm$ .013  & 0.507 $\pm$ .011  & 0.514 $\pm$ .014 
& 0.507 $\pm$ .007  & 0.504 $\pm$ .012  & 0.518 $\pm$ .018  & 0.518 $\pm$ .011  \\ 
mmd-diff-grad 
& \bf 0.531 $\pm$ .009  & \bf 0.529 $\pm$ .006  & 0.532 $\pm$ .008  & \bf 0.566 $\pm$ .006 
& 0.541 $\pm$ .008  & \em 0.528 $\pm$ .011  & 0.535 $\pm$ .009  & \bf 0.554 $\pm$ .016  \\ 
mmd-div-grad 
& 0.525 $\pm$ .011  & \em 0.525 $\pm$ .009  & \em 0.537 $\pm$ .010  & \em 0.563 $\pm$ .011
& \bf 0.532 $\pm$ .013  & 0.53 $\pm$ .012  & \bf 0.54 $\pm$ .011  & \em 0.553 $\pm$ .014  \\ 
nn-comp-greedy
& 0.502 $\pm$ .010  & 0.518 $\pm$ .011  & 0.535 $\pm$ .014  & 0.555 $\pm$ .009 
& 0.515 $\pm$ .01  & 0.528 $\pm$ .011  & 0.528 $\pm$ .009  & 0.543 $\pm$ .009  \\ 
mmd-diff-greedy 
& \em 0.524 $\pm$ .015  & \em 0.520 $\pm$ .013  & 0.521 $\pm$ .013  & 0.537 $\pm$ .010 
& \em 0.521 $\pm$ .017  & \bf 0.529 $\pm$ .016  & \em 0.536 $\pm$ .013  & 0.55 $\pm$ .011  \\ 
mmd-div-greedy 
& 0.517 $\pm$ .012  & 0.515 $\pm$ .010  & 0.519 $\pm$ .012  & 0.532 $\pm$ .011 
& 0.506 $\pm$ .01  & 0.531 $\pm$ .012  & 0.528 $\pm$ .013  & 0.55 $\pm$ .012  \\ 
\hline
\end{tabular}
}
\caption{Classification performance on {\it Gun Control} News dataset. (left) 1-NN, (right) SVM.}
\label{auto_guncontrol}
\vspace{.5cm}
\resizebox{\linewidth}{!}{%
\centering\footnotesize
\begin{tabular}{r|llll|llll}
method & 2 & 4 & 8 & 16 & 2 & 4 & 8 & 16 \\
\hline
kmedoids
& 0.805 $\pm$ .010  & 0.836 $\pm$ .014  & 0.862 $\pm$ .008  & 0.881 $\pm$ .008     
& 0.783 $\pm$ .009  & 0.845 $\pm$ .01  & 0.886 $\pm$ .005  & 0.907 $\pm$ .009  \\ 
kmeans 
& \bf 0.823 $\pm$ .012  & \bf 0.866 $\pm$ .010  & \em 0.888 $\pm$ .006  & 0.909 $\pm$ .009   
& 0.815 $\pm$ .014  & 0.869 $\pm$ .011  & \bf 0.901 $\pm$ .006  & \em 0.924 $\pm$ .009  \\ 
mmd-critic 
& 0.560 $\pm$ .019  & 0.700 $\pm$ .016  & 0.777 $\pm$ .013  & 0.839 $\pm$ .010   
& 0.48 $\pm$ .037  & 0.63 $\pm$ .035  & 0.795 $\pm$ .022  & 0.877 $\pm$ .007  \\ 
mmd-diff-grad 
& \em 0.811 $\pm$ .011  & 0.852 $\pm$ .008  & 0.882 $\pm$ .007  & \em 0.910 $\pm$ .010 
& \em 0.831 $\pm$ .013  & \bf 0.877 $\pm$ .008  & 0.9 $\pm$ .011  & 0.922 $\pm$ .007  \\ 
mmd-div-grad 
&  0.806 $\pm$ .010  & 0.849 $\pm$ .010  & 0.876 $\pm$ .007  & 0.907 $\pm$ .010  
& \bf 0.834 $\pm$ .015  & \em 0.877 $\pm$ .009  & \em 0.901 $\pm$ .011  & 0.919 $\pm$ .008  \\ 
nn-comp-greedy 
& 0.800 $\pm$ .011  & \em 0.859 $\pm$ .010  & \bf 0.890 $\pm$ .009  & \bf 0.914 $\pm$ .010  
& 0.792 $\pm$ .009  & 0.856 $\pm$ .012  & 0.894 $\pm$ .008  & \bf 0.924 $\pm$ .008  \\ 
mmd-diff-greedy 
& 0.783 $\pm$ .011  & 0.835 $\pm$ .009  & 0.871 $\pm$ .009  & 0.898 $\pm$ .010   
& 0.801 $\pm$ .008  & 0.86 $\pm$ .009  & 0.892 $\pm$ .011  & 0.917 $\pm$ .01  \\ 
mmd-div-greedy 
& 0.784 $\pm$ .013  & 0.840 $\pm$ .010  & 0.866 $\pm$ .010  & 0.898 $\pm$ .007   
& 0.798 $\pm$ .008  & 0.866 $\pm$ .012  & 0.895 $\pm$ .01  & 0.92 $\pm$ .009  \\ 
\hline
\end{tabular}
}
\caption{Classification performance on {\it USPS} dataset. (left) 1-NN, (right) SVM.}
\label{auto_usps}
\end{table}

\end{document}